\documentstyle[12pt,epsf,epsfig]{article}     
\def\lsim{\lower0.6ex\vbox{\hbox{$ \buildrel{\textstyle 
<}\over{\sim}\ $}}}
\def\gsim{\lower0.6ex\vbox{\hbox{$ \buildrel{\textstyle 
>}\over{\sim}\ $}}}

\def\beq{\begin{equation}}
\def\eeq{\end{equation}}
\def\beginapjbib{\begingroup \section*{\large \bf References}
         \parskip=.5ex plus 1.0pt
         \def\bibitem{\par \noindent \hangindent\parindent
                \hangafter=1}}
\def\endapjbib{\par \endgroup}
\def\alwaysmath#1{{\ifmmode{#1}\else{$#1$}\fi}}

\def\etal{{\it et al.}~}

\def\eg{{\it e.g.},~}
\def\3he{$^3$He}
\def\4he{$^4$He}
\def\6li{$^6$Li}
\def\7li{$^7$Li}
\def\3h{$^3$H}

\def\la{\mathrel{\mathpalette\fun <}}

\def\fun#1#2{\lower3.6pt\vbox{\baselineskip0pt\lineskip.9pt
  \ialign{$\mathsurround=0pt#1\hfil##\hfil$\crcr#2\crcr\sim\crcr}}}

\begin{document}
\vskip 0.7in
 
\begin{center} 

{\Large{\bf GLOBAL CONSTRAINTS ON KEY}} 
\vskip 0.1in
{\Large{\bf COSMOLOGICAL PARAMETERS}}
 
\vskip 0.4in
{G. Steigman$^{1,2}$, T. P. Walker$^{1,2}$, and A. Zentner$^{1}$}
 
\vskip 0.2in
{\it $^1${Department of Physics,
The Ohio State University, \\ 
Columbus, OH 43210, USA}}\\

\vskip 0.1in
{\it $^2${Department of Astronomy,
The Ohio State University, \\ 
Columbus, OH 43210, USA}}\\



\vskip .75in

{\bf Abstract}\\

\end{center}

Data from Type Ia supernovae, along with X-ray cluster estimates 
of the universal baryon fraction and Big Bang Nucleosynthesis 
(BBN) determinations of the baryon-to-photon ratio, are used to 
provide estimates of several global cosmological parameters at 
epochs near zero redshift.  We show that our estimate of the 
present baryon density is in remarkably good agreement with 
that inferred from BBN at high redshift, provided the primordial 
abundance of deuterium is relatively low and the Universe is 
flat.  We also compare these estimates to the baryon density 
at $z \approx 1100$ as inferred from the CMB angular power 
spectrum.

\newpage
\noindent
\section{Introduction}

In the precision era of cosmology the accuracy in determining 
the key cosmological parameters will be limited by our ability 
to constrain systematic errors and identify observations which 
break the many degeneracies between the global cosmological 
parameters and those related to specific models of inflation 
and/or structure formation.  Redundancy can also provide 
valuable probes of unanticipated systematic uncertainties and 
may serve to separate global from model dependent parameters.  
In particular, although precision measurements of the anisotropy 
of the cosmic microwave background (CMB) promise statistically 
accurate determinations of many cosmological parameters, their 
interpretation may be limited by the extent to which parameter 
degeneracies can be broken and systematic uncertainties can be 
constrained.  As a step in this direction, here we use current 
estimates of a restricted set of {\it global} cosmological 
parameters which are unaffected by ``bias" (in mass versus 
light) and are independent of specific models of structure 
formation and specific theories of inflation to bound a 
variety of the key cosmological parameters.  Such global 
cosmological constraints can then be employed in testing 
models of structure formation and inflation.  

The choice of which observations may provide the best constraints
on the global cosmological parameters is time-dependent and
subjective (see, for example, Steigman, Hata \& Felten 1999 
(SHF); Bahcall \etal 1999).  It is our goal here to minimize 
the observational input while maximizing the predicted output.  
To this end, we first utilize only the magnitude-redshift data 
from the type Ia supernovae (Perlmutter \etal 1997, 1999; Riess 
\etal 1997, 1998; Schmidt \etal 1998; for a recent review and 
extensive references, see Filipenko \& Riess 2000).  In a FRW 
cosmology with an equation of state limited to two components, 
matter with zero pressure ($p =0$) and vacuum energy (or a 
cosmological constant $\Lambda$) with negative pressure, $p 
= -\rho$, the {\it assumption} of flatness, in concert with 
the SNIa data, constrain the matter density ($\Omega_{\rm M}$) 
and the vacuum energy density ($\Omega_{\Lambda}$) (Goobar \& 
Perlmutter 1995).  Armed with $\Omega_{\rm M}$ and 
$\Omega_{\Lambda}$ we determine a variety of the other key 
cosmological parameters such as the deceleration parameter 
($q_0 = {\Omega_{\rm M}\over 2} - \Omega_{\Lambda}$) and the 
dimensionless age of the Universe (H$_0$t$_0$).  Furthermore, 
since a non-BBN, non-CMB constraint on the baryon density is 
of great current interest, we use estimates of the universal 
baryon fraction ($f_{\rm B}$) derived from X-ray observations 
of galaxy clusters to bound the zero-redshift baryon density 
(White \etal 1993 (WNEF); Steigman \& Felten 1995; SHF).  
This independent determination of the baryon density (at 
zero redshift) may be compared to the high redshift estimates 
inferred from BBN (see Olive, Steigman, \& Walker 2000 and 
references therein) and from CMB anisotropy measurements 
(De Bernardis \etal 2000; Hanany \etal 2000; Lange \etal 
2000; Jaffe \etal 2000 and further references therein).  
Next, knowledge of $\Omega_{\rm M}$ and $\Omega_{\rm B}$, 
along with the HST Key Project (Mould \etal 1999) constraint 
on the Hubble parameter permits us to bound other cosmological 
parameters such as the ``shape" parameter $\Gamma$ (Peacock 
\& Dodds 1993; Sugiyama 1995; Peacock 1997; SHF) and to 
estimate the present age of the Universe t$_{0}$. 

We also explore a complementary approach by discarding the 
assumption of flatness ($k = 0$), and instead fixing the 
universal density of baryons from BBN.  This along with 
estimates of the universal baryon fraction from the X-ray 
cluster data permits us to bound the total matter density, 
which may then be combined with the SNIa magnitude-redshift 
data to obtain constraints on the cosmological constant.  
With these constraints on $\Omega_{\rm M}$ and $\Omega_{\Lambda}$ 
we proceed to bound the 3-space curvature ($\Omega_{k} \equiv 1 
- (\Omega_{\rm M} + \Omega_{\Lambda}$)), and the other global 
cosmological parameters such as $q_{0}$, H$_0$t$_0$ (along 
with the age t$_0$), and $\Gamma$. 

\section{Non-BBN, Non-CMB Cosmological Constraints: Maximum 
Returns For Minimum Investment}

Two major groups have mounted systematic investigations of 
the high-redshift SNIa magnitude-redshift relation, the 
``Supernova Cosmology Project" (SCP) of Perlmutter \etal 
(1997, 1999) and the ``High-Z Supernova Search Team" (HZT) 
of Schmidt \etal (1998).  S. Jha and the HZT have kindly 
made available to us the combined likelihoods and it is 
this joint data set we employ in our analysis.  In our
approach the computed quantities are the {\it likelihood 
distributions} for the various cosmological parameters 
(see Figures 2, 3, and 6).  Since none of our resulting
distributions are perfectly gaussian, we report our 
quantitative results in two ways.  We quote results in 
the form $A^{+a_{+}}_{-a_{-}}$, where $A$ is the {\it 
most likely} value and the range from $A - a_{-}$ to 
$A + a_{+}$ defines a 68\% confidence region bounded 
by equiprobable points.  We also quote the full 95\% 
confidence range.

\begin{figure}[ht]
	\centering
	\epsfysize=3.6truein 
\epsfbox{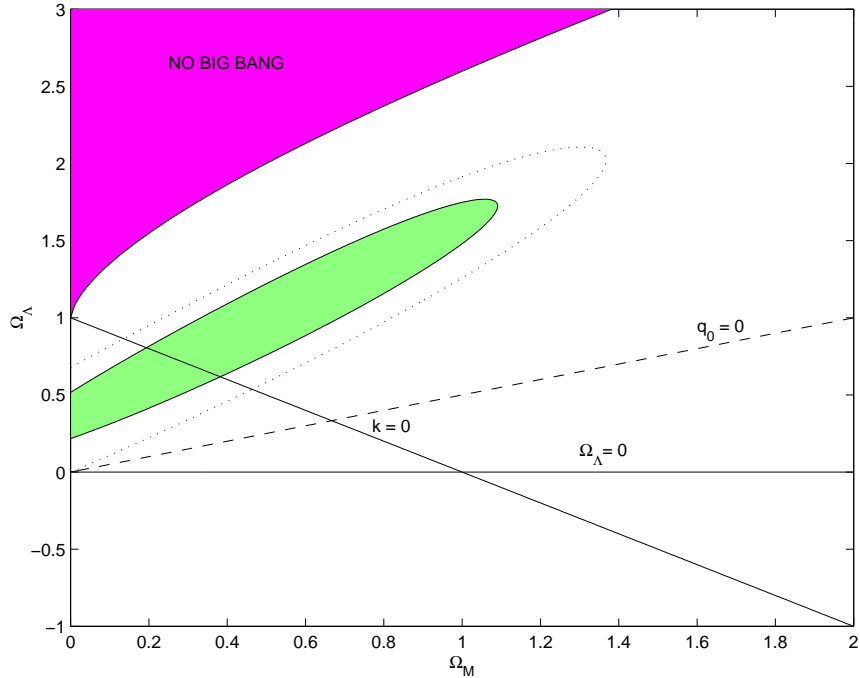}
	\caption{\small{The 68\% (solid) and 95\% (dotted) contours
in the $\Omega_{\Lambda} - \Omega_{\rm M}$ plane allowed
by the magnitude-redshift relation inferred from the 
joint (HZT and SCP) SNIa data.  Closed universes lie in 
the region to the right and above the $k = 0$ line while 
decelerating universes lie below and to the right of the 
$q_{0} = 0$ line (dashed).}}
	\label{fig1}
\end{figure}

As is by now well known, the SNIa data identify a preferred
region in the $\Omega_{\Lambda} - \Omega_{\rm M}$ plane 
(see Figure 1) favoring an {\it accelerating} Universe 
($q_{0} < 0$).  The SNIa contours avoid $\Omega_{\Lambda} 
= 0$, and are cut by the flatness relation ($k = 0$): 
$\Omega_{k} \equiv 1 - \Omega_{\Lambda} + \Omega_{\rm M} 
= 0$.  If flatness is imposed as a constraint, the SNIa 
degeneracy between $\Omega_{\rm M}$ and $\Omega_{\Lambda}$ 
is broken and we find, in agreement with the SCP result of 
Perlmutter \etal (1999) and the HZT result of Filipenko 
\& Riess (2000),
  
\beq
Flat \ (k = 0): \ \ \Omega_{\rm M} \equiv 1 - \Omega_{\Lambda} 
= 0.28^{+0.08}_{-0.07}.
\eeq
The likelihood distribution for $\Omega_{\rm M}$ is shown 
in panel {\it a} of Figure 2.  The corresponding 95\% range, 
$0.15 \leq \Omega_{\rm M} \leq 0.45$, is entirely consistent 
with independent estimates (\eg Cole \etal 1997; Carlberg 
\etal 1997; Bahcall \etal 1999; Weinberg \etal 1999).  

Under the assumption of flatness the deceleration parameter
is $q_{0} = {3\Omega_{\rm M}\over 2} - 1$ which, for 
the above value of $\Omega_{\rm M}$, leads to $q_{0} = 
-0.58^{+0.12}_{-0.10}$.  The likelihood distribution is 
shown in panel {\it b} of Figure 2; the corresponding 95\% 
range is $-0.35 \geq q_{0} \geq -0.77$.  The Universe is 
accelerating.

\begin{figure}[ht]
	\centering
	\epsfysize=3.6truein 
\epsfbox{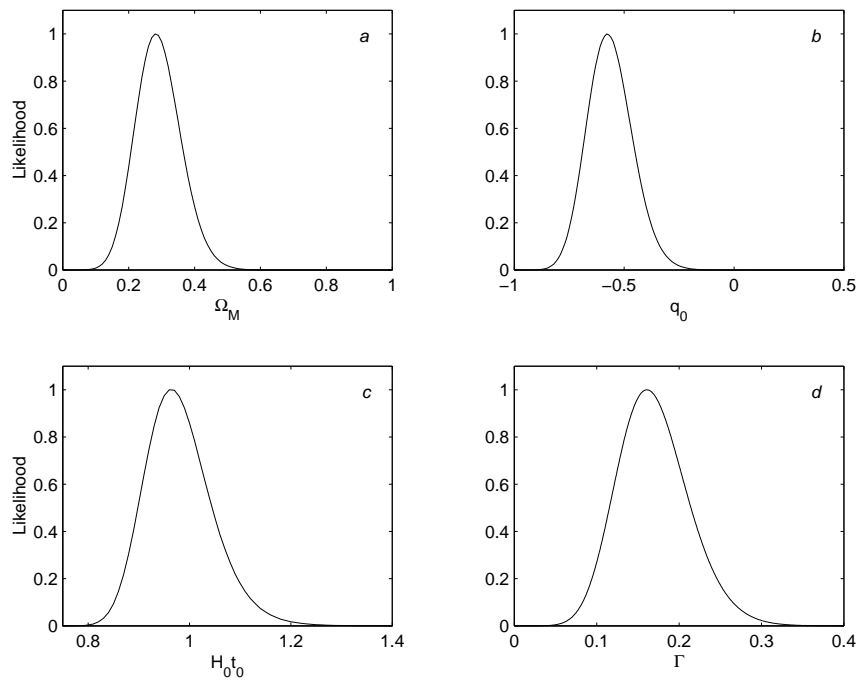}
	\caption{\small{Likelihood distributions, normalized 
to unit maximum, for several global cosmological parameters
as inferred from the SNIa data for a flat universe ($k = 
0$).  Panel {\it a} shows the total (baryonic plus CDM)
density parameter $\Omega_{\rm M}$; panel {\it b} shows
the deceleration parameter, $q_{0}$; the age of the 
Universe in units of the Hubble age, H$_{0}$t$_{0}$, is 
shown in panel {\it c}; the shape parameter $\Gamma$ is 
shown in panel {\it d}.}}
	\label{fig2}
\end{figure}

A flat Universe has a dimensionless age (the age, t$_0$, 
in units of the ``Hubble age", H$_{0}^{-1}$) of H$_{0}$t$_{0} 
= {2\over 3\Omega_{\Lambda}^{1/2}}$sinh$^{-1}
({\Omega_{\Lambda}\over \Omega_{\rm M}})^{1/2}$.  We 
find H$_{0}$t$_{0} = 0.96^{+0.07}_{-0.05}$, in excellent 
agreement with the SCP result H$_{0}$t$_{0} = 0.96^{+0.09}
_{-0.07}$ (Perlmutter \etal 1998) and that from the HZT, 
H$_{0}$t$_{0} = 0.94^{+0.07}_{-0.05}$ (Riess \etal 1998; 
Filipenko \& Riess 2000).  The 95\% range is $0.85 \leq 
~$H$_{0}$t$_{0} \leq 1.13$;  the corresponding likelihood 
distribution is shown in panel {\it c} of Figure 2.  For 
H$_{0} = 71 \pm 6$~kms$^{-1}$Mpc$^{-1}$, (HST Key Project: 
Mould \etal 1999) the age of the Universe is t$_0 = 
13.2^{+1.6}_{-1.3}$~Gyr; the 95\% range, $10.8 \leq 
~$t$_{0}$(Gyr)$~ \leq 16.7$, is in excellent agreement 
with that inferred from globular clusters (Chaboyer 
2000; Chaboyer \& Krauss 2000).

The shape parameter $\Gamma$ (Peacock \& Dodds 1993; 
Sugiyama 1995; Peacock 1997; SHF) depends on $\Omega_
{\rm M}$, $\Omega_{\rm B}$, $h \equiv ~$H$_{0}/100
$kms$^{-1}$Mpc$^{-1}$, and the tilt parameter $n$.  For 
$k = 0$ and $n = 1$ we find $\Gamma = 0.16^{+0.05}_{-0.04}$ 
and a 95\% range of $0.08 \leq \Gamma \leq 0.26$, which 
has considerable overlap with that inferred from observations 
of large scale structure (Fisher, Scharf, \& Lahav 1994; 
Webster \etal 1998; Eisenstein \& Zaldarriaga 1999; 
Efstathiou \& Moody 2000).  The likelihood distribution 
for $\Gamma$ is shown in panel {\it d} of Figure 2.

In summary, the SNIa data and the {\it assumption} of a
flat Universe lead to a set of values for the {\it global}
cosmological parameters $\Omega_{\rm M}$, $\Omega_{\Lambda}$,
and $q_{0}$, along with the large scale structure parameter
$\Gamma$, which are consistent with other observational
data and provide support for a Universe old enough to 
accomodate its oldest stars.

\subsection{X-Ray Cluster Baryon Fraction and the Baryon Density}

Rich clusters of galaxies, the largest collapsed systems 
in the Universe, provide an ideal laboratory for exploring 
the universal baryon fraction (cf. WNEF; SHF).  The X-rays 
observed from the hot intracluster gas may be used to 
estimate both the baryonic mass and the total mass of 
the cluster.  Following WNEF, SHF have written the 
baryon fraction, $f_{\rm B}$, as

\beq
f_{\rm B} = {f_{\rm HG}\over \Upsilon}(1 + {h^{3/2}\over 5.5}),
\eeq
where $f_{\rm HG}$ is the fraction of the total mass in the 
X-ray emitting hot gas, $\Upsilon$ is a baryon enhancement 
factor introduced by WNEF to account for the small offset 
between the {\it universal} and cluster baryon fractions, 
and the last term in eq. 2, taken from WNEF, accounts for 
the cluster baryons which are in stars rather than in the 
hot gas (while ignoring any possible contribution from 
baryonic cluster dark matter).  From a variety of 
hydrodynamical cosmological simulations Frenk \etal 
(1999) find for the offset between the universal and 
cluster baryon fractions: $\Upsilon = 0.92 \pm 0.08$.

SHF used Evrard's (1997) estimate of the hot gas fraction 
and chose $f_{\rm HG}h^{3/2} = 0.060 \pm 0.006$.  More 
recently, Mohr, Mathiesen and Evrard (1999) have summarized 
the observational results from an ensemble of clusters and 
also corrected for the effect of merger driven clumpiness 
(Mathiesen, Evrard \& Mohr 1999) to find $f_{\rm HG}h^{3/2} 
= 0.075$.  Although their quoted formal statistical uncertainty 
is small, systematic errors likely dominate the error budget.  
For example, while the latter estimate has been obtained 
using the ``isothermal beta model" as a total mass estimator, 
Evrard (Private Communication), preferring the virial theorem 
mass estimator, finds $f_{\rm HG}h^{3/2} = 0.056$.  In an 
attempt to account for this uncertainty, here we adopt the 
average of the Evrard (1997) and the Mathiesen, Evrard, \& Mohr 
(1999) determinations, along with a correspondingly generous 
error estimate: $f_{\rm HG}h^{3/2} = 0.066 \pm 0.013$.  
As gravitational lensing observations of X-ray clusters 
improve (see, for example, Tyson, Kochanski \& Dell'Antonio 
1998) this uncertainty should be reduced considerably.  
Observations of the Sunyaev-Zeldovich effect in clusters 
also promise more accurate determinations of the cluster 
hot gas fraction (see, for example, Grego \etal 2000).

The present ratio of baryons to (CMB) photons is parameterized 
by $\eta_{10} \equiv 10^{10}n_{\rm B}/n_{\gamma}$ which, 
for a present CMB temperature of 2.725 K (Mather \etal 1999), 
may be written in terms of the baryon density parameter 
($\Omega_{\rm B}$) and the Hubble parameter ($h$) as,

\beq
\eta_{10} = 274~\Omega_{\rm B}h^{2} = 274~(f_{\rm B}h^{2})
~\Omega_{\rm M}.
\eeq

Combining the above X-ray cluster estimates with the HST 
Key Project determined Hubble parameter (Mould \etal 1999), 
we obtain the distribution for $f_{\rm B}h^{2}$.  We find 
$f_{\rm B}h^{2} = 0.065^{+0.016}_{-0.015}$ and the 95\% range 
is $0.037 \leq f_{\rm B}h^{2} \leq 0.099$.  Convolving this 
with the previously determined distribution (SNIa, $k = 0$) 
for $\Omega_{\rm M}$ we find, 

\beq
Flat \ (k = 0): \ \ \eta_{10} = 4.8^{+1.9}_{-1.5} \ \ \  
(\Omega_{\rm B}h^{2} = 0.018^{+0.007}_{-0.005}).
\eeq

\begin{figure}[ht]
	\centering
	\epsfysize=3.6truein 
\epsfbox{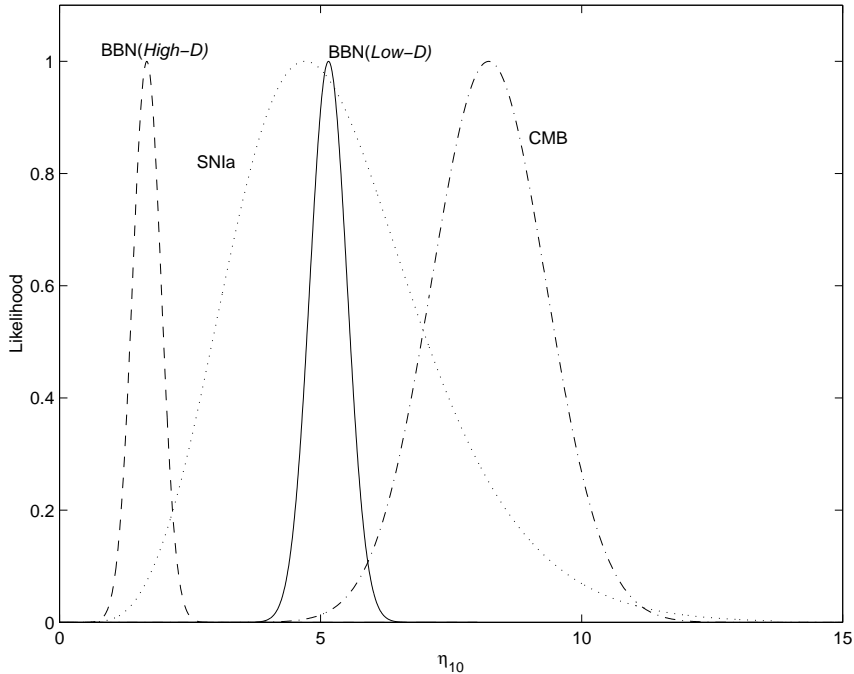}
	\caption{\small{The likelihood distributions, normalized 
to unit maximum, for the baryon-to-photon ratio 
$\eta_{10} = 274\Omega_{\rm B}h^{2}$. The solid 
curve is for the low-D BBN case; the dashed curve 
is for the high-D BBN case (see Sec. 3); the dotted 
curve is for the non-BBN case (SNIa \& $k = 0$); 
the dot-dashed curve is the CMB inferred range for 
$\Omega_{\rm B}h^{2} = 0.032 \pm 0.005$ from Jaffe 
\etal(2000).}}
	\label{fig3}
\end{figure}

The likelihood distribution for $\eta$ is shown in Figure 3.
While the 95\% range, $2.1 \leq \eta_{10} \leq 9.1$ ($0.008 
\leq \Omega_{\rm B}h^{2} \leq 0.033$), is very broad, this 
determination of the baryon density (at zero redshift) is 
independent of, and complementary to, those from BBN ($z 
\approx 10^8$) and the CMB ($z \approx 10^3$).  As will 
become clear in the next section (see, also, Fig.~3), this 
estimate of the baryon density is in excellent agreement 
with that determined from BBN using the ``low'' deuterium 
abundance of Burles \& Tytler (1998a,b; hereafter, BT).  
Although offset from the value of the baryon density 
consistent with BBN and the ``high'' deuterium abundance 
of Webb \etal (1997), given the large errors for both 
determinations there is still considerable overlap.

Fukugita, Hogan, and Peebles (1998) have attempted 
to inventory baryons at $z \approx 0$, finding a range 
(for $h \equiv 0.70$) $0.007 \la \Omega_{\rm B} \la 
0.041$.  Although this range (corresponding to $1.0 \la 
\eta_{10} \la 5.5$) has considerable overlap with ours, 
there is a hint that some dark baryons may have escaped 
the Fukugita, Hogan, \& Peebles (1998) inventory.

Measurements of the ``large $l$'' multipoles of the CMB 
temperature angular power spectrum can also probe the
cosmological parameters.  The recent Boomerang (De Bernardis 
\etal 2000; Lange \etal 2000) and Maxima-1 (Hanany \etal 
2000) CMB anisotropy results point to a ``low'' second 
acoustic peak.  In the attempts to account for a second 
peak which is low relative to the first peak in the CMB 
anisotropy spectrum, there is some degeneracy between 
``global'' cosmological parameters such as the baryon 
density, and model dependent parameters such as ``tilt''.  
Nevertheless, all multi-parameter fits to the Boomerang 
and Maxima data appear to point to a ``high'' baryon density.  
For example, Jaffe \etal (2000) find for their flat model 
($k = 0$) which provides the best fit to the combined 
Boomerang, Maxima and COBE DMR data, $\Omega_{\rm B}h^{2} 
= 0.032 \pm 0.005$.  On the assumption of gaussian errors, 
(almost certainly wrong) the likelihood distribution for 
this result is plotted in Figure 3; note the significant 
overlap between the baryon density determined at present 
from the SNIa data and that from the CMB.  It is a remarkable 
confirmation of the (current) standard model of cosmology 
that these two, independent determinations of the baryon 
density, ``measured'' at vastly different epochs in the 
evolution of the Universe, agree so well.  Given the large 
uncertainties in each determination, the hint of a CMB 
challenge to the zero-redshift baryon density is only 
minor at present. 

\section{Using The BBN-Inferred Baryon Density}

In this section we drop the {\it assumption} of flatness and 
replace it with estimates of the baryon density from BBN.  
Again we use the X-ray cluster inferred baryon fraction for 
an estimate of the universal baryon fraction, but this time we 
combine $f_{\rm B}$ with the BBN measure of $\Omega_{\rm B}h^{2}$
(utilizing the estimates of the primordial abundance of
deuterium) to obtain an estimate of the overall matter 
density (see Figures 4 \& 5):

\beq
\Omega_{\rm M} = {\Omega_{\rm B}\over f_{\rm B}} = 
0.057(1 \pm 0.22)\eta_{10},
\eeq
where we have accounted for the uncertainties in $f_{\rm B}$ 
and $h$ as discussed above.  Armed with the SNIa data and 
this BBN-related estimate of $\Omega_{\rm M}$, we proceed 
to determine the likelihood distributions for the global 
cosmological parameters ($\Omega_{\rm M}$, $\Omega_{\Lambda}$, 
$\Omega_{k}$, $q_{0}$, $\Gamma$, H$_{0}$t$_{0}$).  

\begin{figure}[ht]
	\centering
	\epsfysize=3.6truein 
\epsfbox{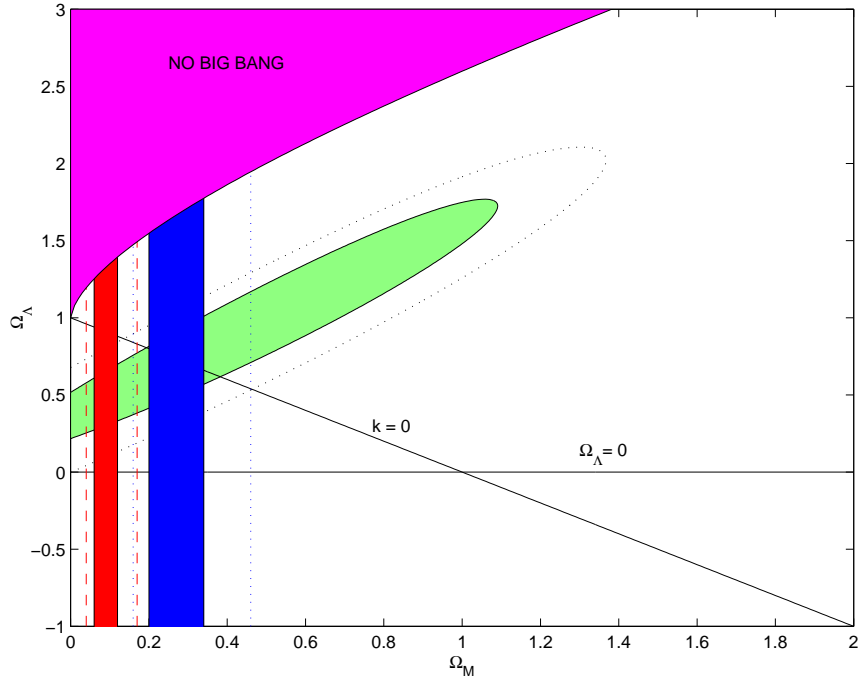}
	\caption{\small{Same as Figure 1 with the 68\% and 95\%
contours for $\Omega_{\rm M}$ inferred from BBN and the 
X-ray cluster baryon fraction.  The high-D, low-$\eta$
constraint bands (at the left) are in red/dashed, while
the low-D, high-$\eta$ bands (at the right) are in 
blue/dotted.}}
	\label{fig4}
\end{figure}

It has long been known that deuterium provides an ideal 
baryometer (Reeves, Audouze, Fowler and Schramm 1976).  
Because deuterium has only been destroyed in the course 
of galactic evolution since the epoch of primordial 
nucleosynthesis (Epstein, Lattimer and Schramm 1976), 
any deuterium, observed anywhere in the Universe, 
provides a {\it lower} limit to its primordial abundance.  
In the high-redshift, low-metallicity environment of the 
QSO absorption-line systems the bulk of the gas is unlikely 
to have been processed through stars, so that the observed 
deuterium abundance is likely to be very nearly the 
primordial value.  For two such systems BT derive a 
statistically accurate estimate for the primordial-D 
abundance: (D/H)$_{\rm P} = 3.4 \pm 0.3 \times 10^{-5}$.  
The $\approx 8$\% observational error is well matched to 
the comparable error in the BBN prediction (Hata \etal 
1996).  Due to the steep dependence of the BBN-predicted 
D-abundance on the baryon-to-photon ratio $\eta$, a 10\% 
uncertainty in (D/H)$_{\rm P}$ results in a baryon 
abundance determined to 6\%.  Combining the observational 
and theoretical errors in quadrature, the BT data lead 
to an estimate of the baryon abundance accurate to 7\%.
The likelihood distribution for $\eta_{10}$ for this low-D 
case is shown in Figure 3 for

\beq
Low-D: ~~ \eta_{10} = 5.1 \pm 0.4 \ \ \ 
(\Omega_{\rm B}h^{2} = 0.019 \pm 0.001).
\eeq
Note the remarkable agreement between this BBN-determined 
baryon density and that obtained in the previous section 
using the X-ray cluster baryon fraction and the SNIa and 
$k = 0$ estimate (see Figure 3 for the corresponding 
likelihood distribution).  The baryon abundance in the 
10 -- 15 Gyr old universe and that inferred for when the 
universe was only a few minutes old are virtually identical 
(as they should be!).   

\begin{figure}[ht]
	\centering
	\epsfysize=3.6truein 
\epsfbox{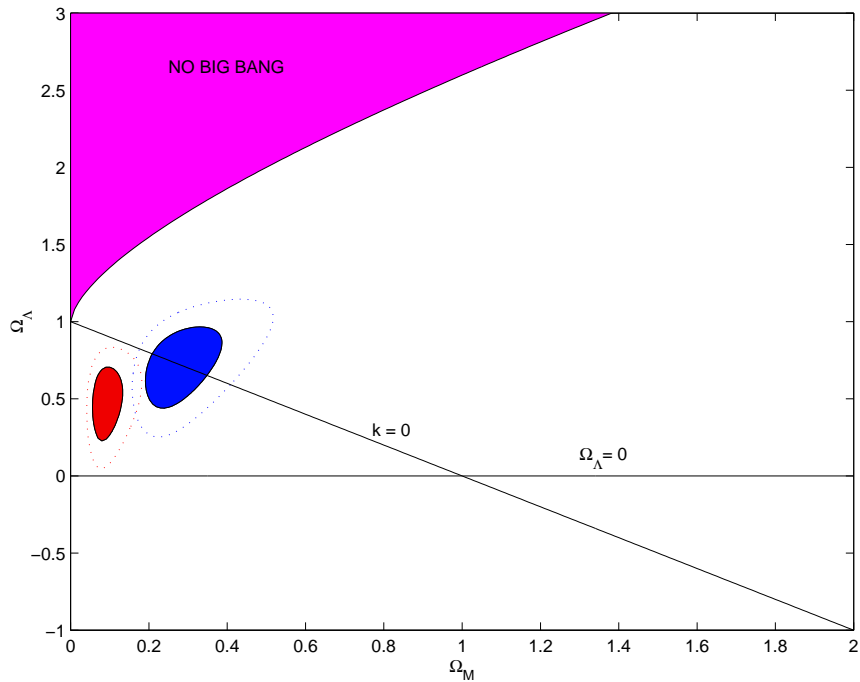}
	\caption{\small{Same as Figure 4, but showing the 68\%
and 95\% contours for the {\it joint} fit of the SNIa 
data with the BBN/X-ray cluster determined values of
$\Omega_{\rm M}$.}}
	\label{fig5}
\end{figure}

In contrast, a higher primordial deuterium abundance, 
(D/H)$_{\rm P} = 20 \pm 10 \times 10^{-5}$ is inferred 
from HST observations of a different, lower redshift 
QSO absorption-line system (Webb \etal 1997).  While 
the high deuterium inferred for this absorption system 
may well be contaminated by a low column density hydrogen 
``interloper'', it could be that this higher abundance is 
primordial.  If it were, then that found by BT would be 
low because the gas in their absorbing systems had been 
cycled through stars where the deuterium was destroyed.  
It would be surprising that the two high redshift systems 
are ``evolved'' while the one lower redshift system 
is not (or, is less evolved).  Clearly more data will 
be crucial in resolving this question.  Here we will 
attempt to use consistency with other cosmological 
parameters as a potential way to discriminate between 
the two possibilities (``low-D'' and ``high-D'').  If 
the Webb \etal (1997) deuterium abundance is adopted, 
the corresponding baryon abundance is both lower and 
somewhat less accurately constrained

\beq
High-D: ~~ \eta_{10} = 1.7 \pm 0.3 \ \ \ 
(\Omega_{\rm B}h^{2} = 0.006 \pm 0.001).
\eeq
From Figure 3 it is clear that the overlap with the
non-BBN estimate from the previous section is poor.

\subsection{Global parameters from low-D}

Here we adopt the BBN-determined baryon density 
corresponding to low-D and we use the X-ray cluster 
baryon fraction to estimate the total mass density.
The corresponding 68\%(95\%) bands are shown 
in Figure 4 and the corresponding contours in 
the $\Omega_{\Lambda} - \Omega_{\rm M}$ plane 
are shown in Figure 5.  Notice that the low-D 
determined contours are completely consistent 
with a flat Universe.  The likelihood distributions 
for several global cosmological parameters are 
shown by the solid curves in panels {\it a -- f} 
of Figure 6.

For the total (baryonic plus CDM) mass density we 
find $\Omega_{\rm M} = 0.26^{+0.09}_{-0.06}$; the
95\% range is $0.16 \leq \Omega_{\rm M} \leq 0.48$.
This is in excellent agreement with the $k = 0$
identified range from the SNIa data and with other, 
independent determinations of the mass density  
(Cole \etal 1997; Weinberg \etal 1999) providing 
support for BBN with low-D. 

In this section we have dropped the restriction
to a flat universe, thereby gaining an independent 
determination of the vacuum density (cosmological 
constant).  The likelihood distribution for 
$\Omega_{\Lambda}$ is shown by the solid curve 
in panel {\it b} of Figure 6.  We find strong 
support for a non-zero value: $\Omega_{\Lambda} 
= 0.74^{+0.17}_{-0.19}$; the 95\% range is $0.35
\leq \Omega_{\Lambda} \leq 1.11$.

Having removed the constraint of flatness we can 
now ask if low-D BBN is consistent with a flat 
Universe.  It is clear from the solid curve in 
panel {\it c} of Figure 6 that a flat universe 
($\Omega_{k} = 0$) is completely consistent with 
the matter density determined from low-D BBN and 
the SNIa magnitude-redshift relation.  Indeed, our 
best fit is for $\Omega_{k} = 0.00^{+0.22}_{-0.24}$
and the 95\% range is $-0.53 \leq \Omega_{k} \leq 
+0.44$.

Given the similarity between the non-BBN parameters 
and those identified by low-D BBN, it is not suprising 
that here, too, we confirm an accelerating universe.  
The likelihood distribution for $q_{0}$ is shown by 
the solid curve in panel {\it d} of Figure 6.  Our 
best fit is for $q_{0} = -0.60^{+0.18}_{-0.16}$ and 
the 95\% range is $-0.24 \geq q_{0} \geq -0.91$.

\begin{figure}[ht]
	\centering
	\epsfysize=3.6truein 
\epsfbox{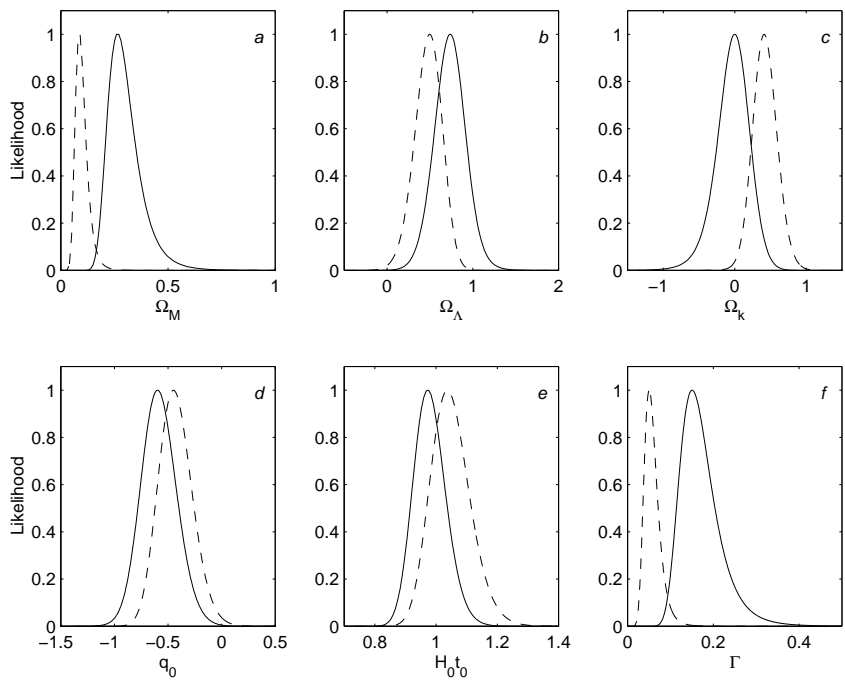}
	\caption{\small{Likelihood distributions for several
global cosmological parameters determined from the
joint fit of the SNIa data, the X-ray cluster 
baryon fraction, and the BBN constraints on the 
baryon density from low-D (solid curves) and high-D 
(dashed curves).  Panel {\it a} shows the total 
mass density parameter $\Omega_{\rm M}$; panel 
{\it b} shows the density parameter associated
with a cosmological constant $\Omega_{\Lambda}$; 
the ``flatness" parameter $\Omega_{k} \equiv 1 -
(\Omega_{\rm M} + \Omega_{\Lambda})$ is shown in
panel {\it c};  the deceleration parameter $q_{0}$
is shown in panel {\it d}; in panel {\it e} is
the dimensionless age H$_{0}$t$_{0}$; the shape
parameter $\Gamma$ is shown in panel {\it f}.}}
	\label{fig6}
\end{figure}

The dimensionless age of the universe, H$_{0}$t$_{0}$,
is shown in panel {\it e} of Figure 6.  For this low-D
universe we find H$_{0}$t$_{0} = 0.97^{+0.07}_{-0.05}$,
with a 95\% range of $0.88 \leq ~$H$_{0}$t$_{0} ~\leq 
1.10$.  Adopting the HST Key project value of H$_{0} =
71 \pm 6$ kms$^{-1}$Mpc$^{-1}$ (Mould \etal 1999), we
find for the age of the Universe t$_{0} = 13.3^{+1.5}_
{-1.3}$ Gyr with a 95\% range extending from 11.0 Gyr
to 16.5 Gyr. 

Finally, we may predict the value of the shape 
parameter corresponding to this low-D case.  The 
likelihood distribution is shown in panel {\it f} 
of Figure 6.  It corresponds to a somewhat low value 
$\Gamma = 0.15^{+0.05}_{-0.04}$, but the 95\% range, 
$0.08 \leq \Gamma \leq 0.28$ is perfectly consistent 
with observational estimates (Peacock \& Dodds 1994; 
Fisher, Scharf \& Lahav 1994; Maddox, Efstathiou \& 
Sutherland 1996; Webster \etal 1998; Eisenstein \& 
Zaldarriaga 1999; Efstathiou \& Moody 2000).

Very recently O'Meara \etal (2000) have reported the
detection of deuterium in another high redshift, low
metallicity QSO absorption system.  The deuterium
abundance they derive, D/H = $2.54 \pm 0.23 \times
10^{-5}$, is some 3$\sigma$ smaller than the previous
mean D/H for the ``low-D" absorbers.  Accounting for
the dispersion among the three, low-D absorbers,
O'Meara \etal suggest a revised primordial deuterium
abundance: D/H = $3.0 \pm 0.4 \times 10^{-5}$, which
for BBN corresponds to a baryon-to-photon ratio of
$\eta_{10} = 5.6 \pm 0.5$, or $\Omega_{\rm B}h^{2} 
= 0.020 \pm 0.002$.  This small shift in the baryon 
density has minimal effect on the quantitative conclusions 
reached above.  For example, while the 95\% ranges 
for $\Omega_{\rm M}$ and $\Omega_{\Lambda}$ shift 
upwards by approximately 0.03 and 0.04 respectively, 
that for $\Omega_{k}$ moves down by $\approx 0.05$, 
and the ranges of $q_{0}$, $\Gamma$, H$_{0}$t$_{0}$, 
and t$_{0}$ are virtually unchanged.

In summary, the low-D BBN estimate of the baryon
density, in combination with the X-ray cluster
estimate of the baryon fraction and the SNIa data
favors a flat ($k = 0$), accelerating ($q_{0} < 0$)
Universe whose age (t$_{0} \geq 11$ Gyr) is consistent
with estimates of the ages of the oldest stars and
with a shape parameter ($0.1 \la \Gamma \la 0.3$)
in agreement with estimates from observations of
large scale structure.

\subsection{Global parameters from high-D}

If, in contrast to the above, the BBN-determined 
baryon density corresponding to high-D (Webb \etal
2000) is used, we find a very low matter (baryons 
plus CDM) density, $\Omega_{\rm M} = 0.09 \pm 
0.03$; the 95\% range is $0.04 \leq \Omega_{\rm M} 
\leq 0.17$.  The corresponding vertical band for 
$\Omega_{\rm M}$ is shown in Figure 4, the related 
$\Omega_{\Lambda} - \Omega_{\rm M}$ contours are 
shown in Figure 5, and the likelihood distribution 
for $\Omega_{\rm M}$ is the dashed curve in panel 
{\it a} of Figure 6.  Such a low value for the 
total mass density is in conflict with independent 
determinations (Bahcall \etal 1999) and with the 
comparison between models of structure formation 
and observations of large scale structure (Cole 
\etal 1997; Weinberg \etal 1999).

The corresponding likelihood distributions of 
the other global cosmological parameters are 
shown in Figure 6.  It is clear from panel 
{\it c} that a flat Universe is very unlikely 
(95\% range: $0.08 \leq \Omega_{k} \leq 0.78$); 
$\Omega_{k} \leq 0$ is excluded at the 99.3\% 
confidence level if the high-D BBN-determined 
baryon density is assumed to provide a fair 
estimate of the present baryon density.

It is clear from Figure 6 that while the high-D, 
BBN determined values of $\Omega_{\Lambda}$, 
$q_{0}$, and H$_{0}$t$_{0}$ differ little 
from their low-D counterparts (at 95\%: $0.15 
\leq \Omega_{\Lambda} \leq 0.79$; $-0.10 \geq 
q_{0} \geq -0.74$; $0.92 \leq ~$H$_{0}$t$_{0} 
\leq 1.19$), the corresponding range for 
$\Gamma$ is very low ($0.01 \leq \Gamma \leq 
0.10$), in conflict with observational estimates 
(Peacock \& Dodds 1994; Fisher, Scharf \& Lahav 
1994; Maddox, Efstathiou \& Sutherland 1996; 
Webster \etal 1998; Eisenstein \& Zaldarriaga 
1999; Efstathiou \& Moody 2000).

In summary, in contrast to the low-D case 
considered above, the high-D estimate of the
baryon density in combination with the X-ray 
cluster estimate of the baryon fraction and 
the SNIa data does not favor a flat ($k = 0$) 
Universe and prefers a low matter density 
($\Omega_{\rm M} \leq 0.17$) and a small value
of the shape parameter ($\Gamma \leq 0.10$) 
which are in marginal conflict with estimates 
from observations of large scale structure.
These latter data argue against high-D and 
favor the low-D estimate of the BBN-predicted 
primordial deuterium abundance.

\section{Discussion}

Data from observations of the SNIa magnitude-redshift 
relation identify a region in the $\Omega_{\Lambda} - 
\Omega_{\rm M}$ plane consistent with a flat universe 
($k = 0$), which disfavors a vanishing cosmological 
constant ($\Omega_{\Lambda} \neq 0$), and points 
towards an accelerating expansion ($q_{0} < 0$) of 
the present universe.  Imposing the {\it assumption} 
of flatness ($k = 0$) breaks the degeneracy between 
$\Omega_{\Lambda}$ and $\Omega_{\rm M}$, identifying 
preferred values for each parameter separately, 
leading to well-defined predictions for a variety 
of other global cosmological parameters (see Figures 
1 \& 2).  In particular, we have supplemented the 
SNIa data with estimates of the cluster baryon 
fraction to relate the total mass density to the 
baryonic mass density, finding a value (determined 
at redshifts $< 1$) consistent with that suggested 
by BBN (fixed at redshifts $> 10^{8}$) for the case 
of a low primordial abundance of deuterium ($\eta_{10} 
\approx 5.1$, $\Omega_{\rm B}h^2 \approx 0.019$).  
Although the CMB-inferred value of the baryon density 
(Jaffe \etal 2000) has a large uncertainty, and 
the result is dependent on non-global cosmological 
parameters, it too is roughly consistent with 
this value (albeit on the high side by some 60\%).  
In contrast, the BBN-determined value of the baryon 
density in the case of a high primordial abundance 
of deuterium ($\eta_{10} \approx 1.7$; $\Omega_{\rm 
B}h^2 \approx 0.006$) is inconsistent with both 
the SNIa -- X-ray cluster result and that from the 
CMB anisotropy measurements.  Our results are 
summarized in Figure 3 where we show the likelihood 
distributions for $\eta_{10}$ for the cases considered 
here, as well as the estimate from the CMB anisotropy 
(Jaffe \etal 2000).  It is clear that there is 
excellent overlap between the non-BBN SNIa range 
(assuming flatness) and the low-D BBN range.  
In contrast, high-D BBN combined with the X-ray 
cluster baryon fraction leads to a very low 
estimate of the total matter density which, when 
combined with the SNIa data, appears inconsistent
with a flat universe. 

In all cases considered here (non-BBN, low-D 
BBN, high-D BBN) the SNIa data strongly favor 
an accelerating Universe ($q_{0} < 0$).  For 
all cases the ``age problem'' has evaporated 
(H$_{0}$t$_{0} \approx 1$, t$_{0} \approx 13 
- 14$ Gyr).  However, although both the non-BBN 
and the low-D BBN results predict a value for the 
shape parameter ($\Gamma \approx 0.2$) consistent 
with observations (Peacock \& Dodds 1994; Fisher, 
Scharf \& Lahav 1994; Maddox, Efstathiou \& 
Sutherland 1996; Webster \etal 1998), the 
high-D prediction ($\Gamma \approx 0.05$) is 
low.  All in all, our analysis supports a flat 
Universe ($\Omega_{k} \approx 0$), presently 
dominated by a non-zero cosmological constant 
($\Omega_{\Lambda} \approx 1 - \Omega_{\rm M} 
\approx 0.7$) or some other form of ``dark 
energy'', and a baryon density consistent 
with that inferred from a comparison of BBN 
and the low value of the deuterium abundance 
found by BT ($\eta_{10} \approx 5$; $\Omega_
{\rm B}h^{2} \approx 0.02$).  While this baryon 
density is roughly a factor of 3 larger than 
that inferred from BBN and the high deuterium 
value of Webb \etal (1997), it is within less 
than a factor of two from that derived from 
the CMB anisotropy data (Jaffe \etal 2000).
Indeed, agreement between the baryon abundance
at epochs in the evolution of the Universe (BBN,
CMB, $z \approx 0$) separated by some 10 -- 14
orders of magnitude in time provides powerful
support for the standard hot big bang cosmological
model.

\vskip 0.5truecm

\noindent {\bf Acknowledgments}

In the preparation of this manuscript we have 
benefitted from the advice and wisdom of many 
colleagues.  In particular, we wish to thank
Saurabh Jha and the High-Z Supernova Search 
Team for providing the combined SCP and HZT 
SNIa data set used in our analysis here and 
A. Balbi for help in utilizing it; Gus Evrard, 
Joe Mohr, and Tony Tyson for helping us understand 
X-ray clusters; Brian Chaboyer for advice on 
the ages of the oldest stars; and Jim Felten, 
Jordi Miralda-Escud\'e, and David Weinberg 
for valuable general discussions.  This 
research is supported by the DOE grant 
to OSU: DE-FG02-91ER-40690. 

\vskip 0.5truecm


\vskip 0.5truecm

\beginapjbib

\bibitem Bahcall, N. A., Ostriker, J. P., Perlmutter, S. \&
Steinhardt, P. J. 1999, Science, 284, 1481

\bibitem Burles, S., \& Tytler, D. 1998a;b, ApJ, 499, 699; 
{\it ibid} 507, 732 (BT)

\bibitem Carlberg, R. G., Yee, H. K. C., \& Ellingson, E. 
1997, ApJ, 478, 462

\bibitem Chaboyer, B. 2000, private communication

\bibitem Chaboyer, B. \& Krauss, L. 2000, in preparation

\bibitem Cole, S., Weinberg, D. H., Frenk, C. S., 
\& Ratra, B. 1997, MNRAS, 289, 37

\bibitem De Bernardis, P. \etal 2000, Nature, 404, 955

\bibitem Efstathiou, G. \& Moody, S. J. 2000, preprint 
(astro-ph/0010478)

\bibitem Eisenstein, D. J. \& Zaldarriaga, M. 1999, 
preprint (astro-ph/9912149)

\bibitem Epstein, R., Lattimer, J., \& Schramm, D. N. 1976,
Nature, 263, 198

\bibitem Evrard, A. E. 1997, MNRAS, 292, 289

\bibitem Filipenko, A. V., \& Riess, A. G. 2000, preprint 
(astro-ph/0008057)

\bibitem Fisher, K. B., Scharf, C. A., \& Lahav, O. 1994,
MNRAS 266, 219
 
\bibitem Fukugita, M., Hogan, C. J., \& Peebles, P. J. E. 1998,
ApJ, 503, 518

\bibitem Frenk, C. S. \etal 1999, ApJ, 525, 554

\bibitem Grego, L. \etal 2000, preprint (astro-ph/0003085)

\bibitem Goobar, A., \& Perlmutter, S. 1995, ApJ, 450, 14

\bibitem Hanany, S. \etal 2000, preprint (astro-ph/0005123)

\bibitem Hata, N., Scherrer, R. J., Steigman, G., Thomas, D.,
\& Walker, T. P. 1996, ApJ, 458, 637

\bibitem Hernquist, L., Katz, N., Weinberg, D. H., \& 
Miralda-Escud\'e, J. 1996, ApJ, 457, L51

\bibitem Jaffe, A. H. \etal 2000, preprint (astro-ph/0007333)

\bibitem Lange, A. \etal 2000, preprint (astro-ph/0005004)

\bibitem Maddox, S. J., Efstathiou, G., \& Sutherland, W. J. 
1996, MNRAS, 283, 1227

\bibitem Mather, J. C. \etal 1999, ApJ 512, 511

\bibitem Mathiesen, B. Evrard, A. E., \& Mohr, J. J. 1999,
ApJ, 520, L21

\bibitem McDonald, P., Miralda-Escud\'e, J., Rauch, M., Sargent, 
W. L. W., Barlow, T., \& Cen R. 2000, preprint (astro-ph/0005533)

\bibitem Mohr, J. J., Mathiesen, B., \& Evrard, A. E. 1999, 
ApJ, 517, 627

\bibitem Mould, J. R. \etal 2000, ApJ, 529, 786 (HST Key Project)

\bibitem Olive, K. A., Steigman, G., \& Walker, T. P. 2000,
Physics Reports, 333-334, 389

\bibitem O'Meara, J. M., Tytler, D., Kirkman, D., Suzuki, N., 
Prochaska, J. X., Lubin, D., \& Wolfe, A. M. 2000, preprint
(astro-ph/0011179)

\bibitem Peacock, J. A. 1997, MNRAS, 284, 885

\bibitem Peacock, J. A., \& Dodds, S. J. 1994, MNRAS, 267, 1020

\bibitem Perlmutter, S. \etal 1997, ApJ, 483, 565

\bibitem Perlmutter, S. \etal 1999, ApJ, 517, 565

\bibitem Rauch, M., Haehnelt, M. G., \& Steinmetz, M. 1997,
ApJ, 481, 601

\bibitem Reeves, H., Audouze, J., Fowler, W. A., \& Schramm, 
D. N. 1976, ApJ, 179, 909

\bibitem Riess, A. G. \etal 1997, AJ, 114, 722

\bibitem Riess, A. G. \etal 1998, AJ, 116, 1009

\bibitem Schmidt, B. P. \etal 1998, ApJ, 507, 46

\bibitem Steigman, G., \& Felten, J. E. 1995, Spa. Sci. Rev., 74, 245

\bibitem Steigman, G., Hata, N., \& Felten, J. E. 1999, ApJ, 510, 564

\bibitem Sugiyama, N. 1995, ApJS, 100, 281

\bibitem Tyson, J. A., Kochanski, G. P., \& Dell'Antonio, I. P. 
1998, ApJ, 498, L107 

\bibitem Webb, J. K., Carswell, R. F., Lanzetta, K. M., Ferlet,
R., Lemoine, M., Vidal-Madjar,, A., \& Bowen, D. V. 1997,
Nature, 388, 250

\bibitem Webster, M., Bridle, S. L., Hobson, M. P., Lasenby, A. N., 
Lahav, O., \& Rocha, G. 1998, ApJ, 509, L65

\bibitem Weinberg, D. H., Miralda-Escud\'e, J., Hernquist, L.,
\& Katz, N. 1997, ApJ, 490, 564

\bibitem Weinberg, D. H., Croft, R. A. C., Hernquist, L., 
Katz, N., \& Pettini, M. 1999, ApJ, 522, 563

\bibitem White, S. D. M., Navarro, J. F., Evrard, A. E., \&
Frenk, C. S. 1993, Nature, 366, 429 (WNEF)

\endapjbib

\end{document}